\documentstyle[pra,aps,amsfonts,twocolumn,epsf]{revtex}  
 
\catcode`\@=11

\def\maketitle2{\par 
\begingroup
\let\cite\@bylinecite
\def\thefootnote{\fnsymbol{footnote}}%
\twocolumn[\@maketitle2\vskip2pc]%
\thispagestyle{plain}\@thanks
\endgroup
\def\thefootnote{\arabic{footnote}}%
\setcounter{footnote}{0}%
\let\maketitle2\relax \let\@maketitle2\relax
\let\@thanks\relax \let\@authoraddress\relax \let\@title\relax
\let\@date\relax \let\thanks\relax \let\@abstract\relax 
\let\@pacs\relax}

\def\abstract#1{\gdef\@abstract{{\par 
\bgroup
\ifdim\prevdepth=-1000pt \prevdepth0pt\fi
\hsize\columnwidth
\dimen0=-\prevdepth \advance\dimen0 by17.5pt \nointerlineskip
\small\vrule width 0pt height\dimen0 \relax}{~~}#1\egroup}}

\def\pacs#1{\gdef\@pacs{{\par 
\bgroup
\hsize\columnwidth \parindent0pt
\ifdim\prevdepth=-1000pt \prevdepth0pt\fi
\dimen0=-\prevdepth \advance\dimen0 by20pt\nointerlineskip
\egroup} PACS numbers:~#1}}

\def\@maketitle2{
\@preprint
\@title
\ifdim\prevdepth=-1000pt \prevdepth0pt\fi
\@authoraddress
\@date
\begin{list}{}{\leftmargin=0.10753\textwidth \rightmargin=\leftmargin
\itemsep=1pc\partopsep=-1pc}
\item\@abstract
\item\@pacs
\end{list}
}

\catcode`\@=12

\begin{document}
\draft

\title{Discrete Wigner functions and the phase space 
representation of quantum teleportation}
 
\author{Juan Pablo Paz}

\address{Departamento de F\'\i sica, ``Juan Jos\'e 
Giambiagi'', Facultad de 
Ciencias Exactas y Naturales, UBA, Pabell\'on 1, Ciudad Universitaria, 
1428 Buenos Aires, Argentina}

\abstract{We present a phase space description of the process of 
quantum teleportation for a system with an $N$ dimensional space of 
states. For this purpose we define a discrete Wigner function which 
is a minor variation of previously existing ones. This function is 
useful to represent composite quantum system in phase space
and to analyze situations where entanglement between 
subsystems is relevant (dimensionality of the space of states
of each subsystem is arbitrary). We also describe how a direct 
tomographic measurement of this Wigner function can be performed.}

\date{\today}  
  
\maketitle2
\narrowtext

\section{Introduction}

Quantum teleportation \cite{Teleport1} 
is a scheme by which an unknown quantum 
state is transported between two parties by only transmitting classical 
information between them. This remarkable task requires also 
that the two parties share the two halfs of a composite system prepared 
in a specific entangled state. In recent years this scheme has
been studied in great detail. Interest in teleportation (whose 
experimental feasibility has also been recently demonstrated in 
simple systems \cite{TeleExperiment}) is motivated by more than
one reason. In fact, teleportation is 
one of the most remarkable processes requiring the use of 
entanglement as essential resource \cite{Teleport1,Chuang-Nielsen}.
Moreover, teleportation can also be conceived as a primitive 
for universal quantum computation \cite{GottChuang}. 
In the seminal work of Bennett et al \cite{Teleport1} the
teleportation of the state of a qbit (a two level system) was
described. Shortly afterwards, Vaidman \cite{Vaidman}
showed how to generalize this method to include 
systems with a space of states
of arbitrary dimensionality. In this way, the idea of teleporting
states of systems with continuous variables was first proposed. 
A concrete analysis of the teleportation of the state of a 
continuous system (the electromagnetic field) was first discussed 
by Braunstein and Kimble \cite{Braunstein}. In their work, these 
authors proposed (and later performed \cite{Kimbleexp}) 
interesting experiments to accomplish teleportation of continuous variables. 
For this case, it was natural to describe the whole procedure 
in terms of phase space distributions \cite{Wigner}. 
However, this is not the case for systems with a finite 
dimensional Hilbert space, where the use of the phase space 
representation is not so common. The description of the 
usual teleportation protocol in phase space has been recently 
presented by Koniorczyk, Buzek and Jansky \cite{Buzek}
using the discrete version of Wigner functions originally introduced 
by Wooters \cite{Wooters}. This approach, as mentioned in \cite{Buzek}, 
can only be used when the dimension of the Hilbert space of the system 
to be teleported is a prime number. 
In this paper we extend the results presented in \cite{Buzek} 
to the case where the space of states has arbitrary dimensionality. 
For this purpose we use a different
definition for the discrete Wigner function that turns out
to be very convenient to analyze situations where 
entanglement between subsystems of arbitrary dimensionality 
is an important issue. 

Several methods exist to represent  
the quantum state of a system with an $N$ dimensional Hilbert 
space in phase space. 
As mentioned above, Wooters introduced a discrete version
of the Wigner function that has all the desired properties only 
when $N$ is a prime number\cite{Wooters}. 
His phase space is an $N\times N$ grid 
(if $N$ is prime) and a Cartesian product of such spaces 
corresponding to prime factors of $N$ in the most general case. 
On the other hand, a different approach to define a Wigner function 
for a system with an $N$ dimensional Hilbert space was introduced
by Leonhardt \cite{Leonhardt} that rediscovered results previously 
used by Hannay and Berry \cite{Hannay} and others. 
This method, that has the 
property of being well defined for arbitrary values of $N$, was
used in several contexts \cite{Rivas,Bouzouina}
and recently applied to analyze the 
phase space representation of quantum computers and algorithms
\cite{WignerUS,MPS01}. In this case, the phase space is constructed
as a grid of $2N\times 2N$ points where the state is represented
in a redundant manner (only $N\times N$ of them are truly independent). 
In this paper, we use a hybrid approach allowing us to capture
the most useful features of both Wooters and Leonhardt methods. Thus, to 
represent a quantum state of a bipartite system we use, following Wooters, 
a phase space which is a Cartesian product of two grids. Each one of these
grids has, as it does in Leonhardt approach, $2N_i\times 2N_i$ points 
(where $N_i$ is the dimensionality of the Hilbert space of 
the $i$--th subsystem). 

The paper is organized as follows: 
In Section II we first review the usual approach to define Wigner
functions for systems with an $N$  dimensional Hilbert space. Then 
we introduce a convenient generalization that can be adopted in order
to analyze bipartite (or multi--partite) systems. We discuss some
general properties of the Wigner function and analyze the phase 
space representation of a family of entangled states (generalized 
Bell--EPR states). In Section 3 we show how
to describe teleportation of the quantum state of a system with 
an $N$ dimensional Hilbert space using the Wigner 
function. In Section IV we describe the procedure by which a 
direct measurement of the Wigner function of a composite system
can be done. In Section V we summarize our conclusions. 

\section{Phase space representation of composite quantum 
systems}

We describe here the formalism of Wigner functions 
to represent a composite quantum system  in phase space. For 
simplicity, we consider the composite system to be formed by 
subsystems each one of which has an $N$ dimensional Hilbert space
We first describe the properties 
of discrete Wigner functions for one of the 
subsystems and later discuss the phase space representation of the 
composite system. It is clear that 
one can always split a system 
into subsystems in many ways. For example, 
when $N$ is a composite number one can choose to separate each 
subsystem into even smaller subsystems each one of which has 
a space of states with 
the dimensionality of the prime factors of $N$. Adopting this 
description is simply a matter of physical convenience. Here, we 
assume that the $N$ 
dimensional subsystems are the relevant elementary 
components and that, in some physically interesting 
regime, the entanglement between them can be manipulated.  

\subsection{Discrete Wigner functions}

To represent the quantum state in phase space 
we must first define the notions of position and momentum. 
To do this, we introduce a basis of the Hilbert space 
$B_x=\{|n\rangle,n=0,\ldots,N-1\}$ which we arbitrarily 
interpret as the position basis (with periodic boundary 
conditions: $|n+N\rangle=|n\rangle$). Given the position 
basis $B_x$, we introduce the conjugate momentum 
basis $B_p=\{|k\rangle,k=0,\ldots,N-1\}$ by means of the discrete Fourier 
transform. The states of $B_p$ can be obtained from those of $B_x$ 
as
\begin{equation}
|k\rangle={1\over\sqrt{N}}\sum_n \exp(i 2 \pi nk/N)|n\rangle. 
\label{DFT}
\end{equation}
As in the continuous case, position and momentum are
related by the discrete Fourier transform. 
The correct semi-classical limit corresponds to the 
large $N$ limit since the dimensionality of the Hilbert space is related
to an effective Planck constant
as $N=1/2\pi\hbar$. 

Displacement operators in position and momentum, 
denoted as $\hat U$ and $\hat V$, are defined as:
\cite{Schwinger}: 
\begin{eqnarray}
\hat U^m |n\rangle &=& |n+m\rangle,\ \hat U^m |k\rangle = 
\exp(-2\pi i mk/N)|k\rangle,\nonumber\\
\hat V^m |k\rangle &=&|k+m\rangle,\ \hat V^m|n\rangle = 
\exp (i 2 \pi mn/N)|n\rangle.\label{uandv}
\end{eqnarray}
Commutation relations between $\hat U$ and $\hat V$  
directly generalize the ones corresponding to 
finite translations in the continuous case:
\begin{equation}
\label{fact4}
\hat V^p \hat U^q=\hat U^q \hat V^p \exp(i 2 \pi pq/N).
\end{equation}
A reflection operator $\hat R$ can also be defined 
as the one acting in the position  
basis as $\hat R|n\rangle = |-n\rangle$ 
(again, this operation is to be understood mod $N$). 
$\hat R$ is related to the Fourier transform $U_{FT}$ 
(where $\langle n'| U_{FT}| n\rangle = \exp(i2\pi n n'/N)$)
since $\hat R=U_{FT}^2$.

To represent the state in phase space we use the Wigner
function defined as the following expectation 
value \cite{Leonhardt,MPS01}:
\begin{equation}
\label{wigdef}
W(\alpha)=\mbox{Tr}(\hat A(\alpha)\hat \rho)
\end{equation}
where $\alpha$ denotes a phase space point ($\alpha=(q,p)$) 
and $\hat A(\alpha)$ are the so--called ``phase space point 
operators'' defined in terms of displacements 
and reflections as \cite{MPS01,Leonhardt} 
\begin{equation}
\hat A(\alpha)= 
{1 \over {2N}} \hat U^q \hat R \hat V^{-p} \exp(i \pi pq/N).
\label{disphase2}
\end{equation}
It is important to mention that, in order for $W(\alpha)$
to have all the desired properties  
the phase space should be a grid of $2N\times 2N$ 
points which for the rest of the paper will be denoted as $G_{2N}$ 
(i.e., $G_{2N}$ is the set of points $\alpha=(q,p)$ where $q$ 
and $p$ take values between $0$ and $2N-1$). It will also be 
useful to denote the first $N\times N$ sub-grid as $G_N$ 
(i.e., $G_N$ is the set of points $\alpha=(q,p)$ where 
$q$ and $p$ take values between $0$ and $N-1$). 
 
The Wigner function (\ref{wigdef})  
obeys three defining properties: (P1) It is real valued, 
which is a consequence of the fact that the operators 
$\hat A(\alpha)$ are hermitian by construction. (P2) The Wigner 
function can be used to compute expectation values between states
as 
\begin{equation}
\mbox{Tr}[\rho_A \rho_B]=N \sum_{\alpha\in G_{2N}} 
W_A(\alpha) W_B(\alpha). 
\end{equation}
This is a consequence of the completeness of the set of 
operators $\hat A(\alpha)$, that satisfy
\begin{equation}
\mbox{Tr}[\hat A(\alpha) \hat A(\alpha') ] = 
{1\over {4N}} \, \delta_N(q'-q) \delta_N(p'-p) 
\end{equation}
where $\alpha, \alpha'\in G_N$ and 
$\delta_N(q) \equiv {1\over N}\sum_{n=0}^{N-1} \mbox{e}^{2 \pi i q/N}$ 
is the periodic delta function (which is zero unless 
$q=0 \ \mbox{mod} \ N$). 

As $\hat A(\alpha)$ 
form a complete orthonormal basis of the space of operators, one 
can expand the density matrix $\rho$ in this basis  
and show that the Wigner function are nothing but the coefficients
of such expansion. Thus: 
\begin{eqnarray}
\hat \rho &=& 4N \sum_{\alpha\in G_N} W(\alpha) \hat A(\alpha) 
\label{rhofromwN}\\
&=& N  \sum_{\tilde\alpha\in G_{2N}} W(\tilde\alpha) \hat 
A(\tilde\alpha). 
\label{rhofromw2N}
\end{eqnarray}
The last expression, where the sum is over $\alpha\in G_{2N}$, 
can be obtained from (\ref{rhofromwN}) by noticing that 
the contribution of each of the four $N\times N$ sub-grids are 
identical. This can be shown by using the fact that
(for $\sigma_q,\sigma_p=0,1$):
\begin{equation}
\label{Arelations}
\hat A(q+\sigma_q N, p+\sigma_p N)=
\hat A(q,p) \, (-1)^{\sigma_p q + \sigma_q p + \sigma_q \sigma_p N}. 
\end{equation}

Finally, $W(\alpha)$ satisfies a third crucial property. Consider
a line $L$ in the grid $G_{2N}$ (a line $L$ is the set of 
all points $\alpha=(q,p)$ such that $ap-bq=c$ for given integers $a$, 
$b$ and $c$). Then the sum of $W(\alpha)$ over all points 
lying on any line $L$ is always positive. This  
property (P3) is a consequence of the following fact:
Adding all phase space operators over a line $L$ defined by the 
equation $ap-bq=c$ one obtains a projection operator. Thus 
$\hat A_L=\sum_{\alpha\in L} \hat A(\alpha)$ is a projector 
onto an eigenspace of the phase space translation operator 
\begin{equation}
\hat T(a,b)= U^a V^b \exp(i\pi ab/N)
\end{equation} 
with eigenvalue $\exp(i\pi c/N)$.
(If $T(a,b)$ does not have $\exp(i\pi c/N)$ as one of its 
eigenvalues the projector $\hat A_L$ vanishes). The simplest example of the 
use of this property is the following: consider the horizontal lines 
$p=c$. The sum of $W(\alpha)$ over these lines vanishes
if $c$ is odd (because $T(1,0)=U$ has eigenvalues $\exp(i\pi k/N)$ 
when $k$ is an even integer). On the other hand, when $c$ is even, 
the sum of $W(\alpha)$ is equal to the probability of measuring 
a momentum equal to $c/2$. Thus, 
$\sum_{q} W(q,p)= \langle p/2|\hat\rho|p/2\rangle$ if $p$ is even
(and zero otherwise).

Let us summarize the results presented so far: The Wigner function 
for systems with an $N$ dimensional Hilbert space is defined
in (\ref{wigdef}) as the expectation value of the phase space operator
$\hat A(\alpha)$ given in (\ref{disphase2}). This definition is 
such that $W(\alpha)$ is real, it can be used to compute
inner products between states and it gives all the correct 
marginal distributions when added over any line in the phase 
space, which is a grid $G_{2N}$ with $2N\times 2N$ 
points. The size of the phase space grid is important to obtain 
a Wigner function with all the desired properties. The values of  
$W(\alpha)$  on the sub-grid $G_N$ are enough to reconstruct 
the rest of the phase space (since the set $\hat A(\alpha)$ is 
complete when $\alpha$ belong to the grid $G_N$). However, the 
redundancy introduced by the doubling of the number of sites in 
$q$ and $p$ is essential when one imposes the condition that 
all the marginal distributions should be obtained from the Wigner
function. 

As an example, we show here the Wigner function of a position 
eigenstate $\hat\rho = |q_0\rangle\langle q_0|$ (see 
\cite{MPS01} for other interesting examples):
\begin{equation}
W(q,p)= {1 \over {2N}} \delta_N(q-2 q_0) \ (-1)^{p(q-2q_0)_N}.
\label{wigcompstate}
\end{equation}
Here, $z_N$ denotes $z$ modulo $N$. $W(\alpha)$ is non--zero only on
two vertical strips located at $q=2q_0$ modulo $N$. When $q=2q_0$ we have
$W(2q_0,p)=1/2N$ while for $q=2q_0 \pm N$ the modulating factor 
$(-1)^p$ produces oscillations. They can be interpreted as 
the interference between the $q=2q_0$ strip and its mirror image 
induced by the periodic boundary conditions. 
The fact that $W(\alpha)$ becomes negative in this interference strip 
is essential to recover the correct marginal distributions. Adding
the values of $W(q,p)$ along a vertical line gives the probability of
measuring $q/2$ which should be $1$ for $q=2q_0$ and zero otherwise.

\subsection{Wigner function for composite systems}

We now consider a composite system with Hilbert space 
${\cal H}^{(1,2)}={\cal H}^{(1)}\otimes {\cal H}^{(2)}$
(for simplicity we assume that the dimension of 
both spaces ${\cal H}^{(i)}$ is the same but the method 
can be generalized if this is not the case). 
To represent the states of this composite system in phase space
we generalize the approach described in the previous subsection 
in an obvious way: We consider the phase space for the 
composite system as the Cartesian product of the ones for  
the subsystems (as in the classical case) and use the phase space 
point operators defined as 
\begin{equation}
\label{aalpha12}
\hat A(\alpha_1,\alpha_2)=\hat A(\alpha_1)\otimes \hat A(\alpha_2). 
\end{equation}
The set $\{\hat A(\alpha_1,\alpha_2)$, with $\alpha_i\in G_N\}$, is a 
complete orthonormal basis of the space of operators on ${\cal H}^{(1,2)}$
since:
\begin{eqnarray}
{\mbox{Tr}}_{1,2}(\hat A(\alpha_1,\alpha_2) \hat A(\alpha'_1,\alpha'_2)) 
= {1\over {(4N)^2}} &\,& \delta_N(\alpha_1-\alpha'_1)\nonumber\\ 
&\times&\delta_N(\alpha_2-\alpha'_2), 
\end{eqnarray}
where $\alpha_i,\alpha'_i\in G_N$ and 
$\delta_N(\alpha)=\delta_N(q)\delta_N(p)$. 

The Wigner function for the composite system is defined 
as the expectation value of 
the above operators:
\begin{equation}
W(\alpha_1,\alpha_2)={\mbox{Tr}}(\hat A(\alpha_1,\alpha_2) \rho)
\end{equation}     
This function has the same properties than the one 
for each subsystem. In fact, the three properties
(P1--P3) generalize trivially to this case. Reality (P1) is once again
an obvious consequence of the hermitian nature of the phase space 
point operators. Completeness of the operators $\hat A(\alpha_1,\alpha_2)$
enable us to expand the total density matrix in this basis and write:
\begin{eqnarray}
\rho&=&(4N)^2\sum_{\alpha_1,\alpha_2\in G_N} W(\alpha_1,\alpha_2) 
\hat A(\alpha_1,\alpha_2),\nonumber\\
&=&N^2\sum_{\alpha_1,\alpha_2\in G_{2N}} 
W(\alpha_1,\alpha_2) \hat A(\alpha_1,\alpha_2).
\end{eqnarray} 
Once again, the first line (where both $\alpha_1,\alpha_2\in G_N$) 
can be transformed into the second line where the sum 
can be extended to the grid $G_{2N}$ by noticing that 
the contribution over sub-grids are identical (due to 
the relations obtained by using (\ref{Arelations})).
Using this, it is simple to show that inner products between 
two states of the composite system can be computed from the 
Wigner functions as (P2):
\begin{equation}
\mbox{Tr}(\rho_A \rho_B)=N^2 \sum_{\alpha_1,\alpha_2\in G_{2N}} 
W_A(\alpha_1,\alpha_2)W_B(\alpha_1,\alpha_2).
\end{equation}

The third property (P3) is valid as well but it is worth 
discussing it with more detail. The Wigner function 
turns out to be positive when added over any ``slice'' 
of the total phase space. A ``slice'' in phase space can 
be defined (following Wooters \cite{Wooters}) as a 
set of lines $\{L_1,L_2\}$ (one line for each subsystem). These
sets are called slices since in the continuous limit, 
the set of phase space points satisfying the equations
$a_1p_1-b_1 q_1=c_1$ (that defines $L_1$) and 
$a_2p_2-b_2 q_2=c_2$ (defining $L_2$) form a two dimensional
manifold immersed in the four dimensional phase space. The fact 
that the Wigner function is positive when added over all 
points belonging to the slice $\{L_1,L_2\}$ is obvious: Thus, 
adding $\hat A(\alpha_1,\alpha_2)$ over all points 
where $\alpha_1$ and $\alpha_2$ 
respectively belong to the lines $L_1$ and $L_2$   
one always obtains a tensor product of two projectors. These 
operators project onto the eigenspaces of the operators  
\begin{eqnarray}
T_1&=&T(a_1,b_1)\otimes \mbox{I},\nonumber\\
T_2&=&\mbox{I}\otimes T(a_2,b_2)\nonumber
\end{eqnarray}
with eigenvalue $\exp(i\pi c_1/N)$ and  $\exp(i\pi c_2/N)$ 
respectively. We omit the proof of this property, that can be done by 
copying the one presented in \cite{MPS01} (see also below for a 
simple proof of a related property). 
Slices defined as above, by picking a line for each subsystem, 
will be denoted ``separable slices''. 
As we have just seen, Wigner functions when added over separable
slices can be used to compute probabilities for the outcomes of 
measurements of separable observables. These are properties 
that are measured by independent experiments on the two separate
subsystems. 

It is possible, however, to consider more general non--separable 
slices in the phase space. As we will show below, by adding phase 
space operators on these non--separable slices we will obtain projectors 
onto entangled states. For this we should use a general 
kind of manifold $L_{1,2}$ that 
can be defined as the set of points $(\alpha_1,\alpha_2)$ 
satisfying the condition 
$a_1p_1-b_1 q_1+a_2p_2-b_2 q_2=c_{12}$ (notice that calling $L_{12}$ a
line can be misleading: in the continuous limit it is a three 
dimensional manifold). It is simple to show that 
by adding phase space operators over all points belonging to 
$L_{1,2}$ one also obtains a projection operator. It is rather
instructive to see this proof:
For this, we just have to use the fact that the 
Fourier transform of the subsystem's phase space point 
operators $\hat A(\alpha_i)$
is a translation \cite{MPS01}: 
\begin{equation}
T(a,b)=\sum_{q,p=0}^{2N-1} \, \hat A(q,p) \
\exp(-i {2\pi\over{2N}}(ap-bq)). 
\label{ftofa}
\end{equation}
Using this, we can compute the sum of $\hat A(\alpha_1,\alpha_2)$
over $L_{1,2}$ as follows:
\begin{eqnarray}
\hat A_{L_{1,2}}&\equiv&\sum_{(\alpha_1,\alpha_2)\in S_{1,2}} 
\hat A(\alpha_1,\alpha_2)
\nonumber\\
&=&\sum_{\alpha_1,\alpha_2\in G_{2N}} \hat A(\alpha_1)\otimes
\hat A(\alpha_2)\nonumber\\
&\,&\qquad\qquad\times
\delta_{2N}(a_1p_1-b_1 q_1+a_2p_2-b_2 q_2-c_{12})\nonumber\\
&=&{1\over 2N} \sum_{\lambda=0}^{2N-1} T^\lambda(a_1,b_1)\otimes
T^\lambda(a_2,b_2)\, \mbox{e}^{i{\pi\over N} \lambda\, c_{12}}, 
\end{eqnarray}
where to obtain the last line we represented the delta--function
as a sum of exponentials and used (\ref{ftofa}). As the 
translation operators are unitary and cyclic one can 
always express them in terms of their eigenstates and eigenvalues as
\begin{equation}
T(a_i,b_i)=\sum_{n=0}^{N-1}|\phi_{i,n}
\rangle\langle \phi_{i,n}| \exp(-i2\pi n/N). 
\end{equation}
Using this equation one finally obtains:
\begin{eqnarray}
\hat A_{L_{1,2}}=\sum_{n,m} |\phi_{1,n}
\rangle\langle \phi_{1,n}|&\otimes& |\phi_{2,m}
\rangle\langle \phi_{2,m}|\nonumber\\
&\times&\delta_{2N}(n+m-c_{1,2}/2).\label{AL12} 
\end{eqnarray}
This explicitly shows that $A_{L_{1,2}}$ is a projector 
onto the eigenspaces of a collective operator of the bipartite 
system. Thus, (\ref{AL12}) projects onto eigenspaces of 
\begin{equation}
\hat T_{1,2}=\hat T(a_1,b_1)\otimes \hat T(a_2,b_2)
\end{equation}
with eigenvalue  
$\exp(i\pi c_{1,2}/N)$. It is clear that, when $N$ is even 
the projector is non-vanishing only if $c_{1,2}$ is even. 
The most important conclusion is that, generically, the sum of 
phase space point operators over non--separable manifold $L_{12}$ will 
correspond to a projector over an entangled state. In the previous
case $\hat T_{1,2}$ is a collective operator generating 
simultaneous phase space translations of both systems. 
In the continuous limit it is clear that this operator is 
generated by linear combinations of the momenta and coordinates
of the two subsystems. 

In general, a manifold $L_{12}$ will not be associated with 
a one dimensional projector. For example, the manifold 
$L_{+,p_\beta}$ defined by the equation $p_1+p_2=2p_\beta$ corresponds 
to an $N$ dimensional subspace. The same is true for the manifold 
$L_{-,q_\beta}$ defined as the set of points satisfying $q_1-q_2=2q_\beta$. 
The intersection between these two sets will be denoted as 
$L_{\beta}$ (defined as
the set of points belonging to both $L_{+,p_\beta}$ and 
$L_{-,q_\beta}$) and corresponds to a one dimensional projector 
over an entangled state (see below). 

Wigner functions of separable states 
have very different features than those of  
entangled states: In fact, if the density matrix is a tensor
product $\rho=\rho_1\otimes\rho_2$ 
then $W(\alpha_1,\alpha_2)$ is a product of the form  
$W(\alpha_1,\alpha_2)=W_1(\alpha_1)W_2(\alpha_2)$. More generally, if
the state is separable (i.e., the density matrix is a convex sum
of tensor products) then the Wigner function is a convex sum of 
products of the above form. For entangled states (states which are
not separable) this is not the case as will be explicitly seen 
below. It is also useful to notice that reduced Wigner functions 
can be computed for one subsystem 
by adding the total Wigner function over the 
complementary half of the phase space. Thus, summing $W(\alpha_1,\alpha_2)$
over $\alpha_2$ is equivalent to taking the partial trace 
over the second subsystem since
\begin{eqnarray}
W_1(\alpha_1)&=&\sum_{\alpha_2\in G_{2N}} W(\alpha_1,\alpha_2)\nonumber\\
&=&{\mbox{Tr}}_1( \hat A(\alpha_1)\rho_1),
\end{eqnarray}
where $\rho_1$ is the reduced density matrix of the first system
obtained from the total density matrix as $\rho_1={\mbox{Tr}}_2\rho$.
Finally, it is also useful to notice other properties of the composite Wigner
function when we sum it over half of the phase space. For example, 
if $\rho_A$ and $\rho_B$ are two states of the composite system, 
then
\begin{eqnarray}
F(\alpha_1,\alpha'_1)&\equiv&\sum_{\alpha_2\in G_N} W_A(\alpha_1,\alpha_2) 
W_B(\alpha'_1,\alpha_2)\nonumber\\
&=&{1\over 4N} {\mbox{Tr}}_2( {\mbox{Tr}}_1(\rho_A \hat A(\alpha_1))
{\mbox{Tr}}_1(\rho_B \hat A(\alpha'_1)),\label{lastrelation}
\end{eqnarray}
an equation that will useful later. 

\subsection{Wigner functions for Bell states}

Let us first introduce a complete basis of entangled states 
(generalized Bell states \cite{Buzek}). First we define the 
state $|\Theta_0\rangle$ as
\begin{equation}
|\Theta_0\rangle={1\over\sqrt N}\sum_{n=0}^{N-1} 
|n\rangle_1\otimes |n\rangle_2. 
\end{equation}
This pure state for the composite system is maximally entangled since
the reduced density matrix of each subsystem is proportional to the 
identity. A complete basis of entangled states can be defined
from $|\Theta_0\rangle$ as follows:
\begin{equation}
|\Theta_\beta\rangle =
{1\over \sqrt N}\sum_{n=0}^{N-1} {\mbox{e}}^{i2\pi p_\beta n/N} 
|n\rangle_1\otimes |n-q_\beta\rangle_2.\label{bellstates}
\end{equation}
These states can all be obtained from $|\Theta_0\rangle$ by using 
one of the following equivalent expression:
\begin{eqnarray}
|\Theta_\beta\rangle &=& 
T_1(q_\beta,p_\beta)\, {\mbox{e}}^{i\pi q_\beta p_\beta/N}
\otimes {\mbox{I}}_2 |\Theta_0\rangle\nonumber\\
&=&V_1^{p_\beta}\otimes U_2^{-q_\beta}|\Theta_0\rangle\nonumber\\
\end{eqnarray}
Above, the notation $\beta=(q_\beta,p_\beta)$ was used. 
When $\beta\in G_N$ these states form a complete orthonormal 
basis of the Hilbert space ${\cal H}^{(1,2)}$. In fact, these $N^2$ 
vectors satisfy 
\begin{equation}
\langle \Theta_\beta|\Theta_{\beta'}\rangle=\delta_N(\beta-\beta').
\end{equation}
It is interesting to notice that these  
are the common eigenstates of the following complete set of 
commuting operators: 
\begin{eqnarray}
U_+&=&U_1\otimes U_2,\nonumber\\
V_-&=&V_1\otimes V_2^\dagger.\label{u+v-}
\end{eqnarray}
In fact, $V_-$ and $U_+$ commute and that Bell states 
(\ref{bellstates}) obey the following identities:
\begin{eqnarray} 
U_+|\Theta_\beta\rangle&=&\exp(-i2\pi p_\beta/N) |\Theta_\beta\rangle,
\nonumber\\
V_-|\Theta_\beta\rangle&=&\exp(i2\pi q_\beta/N) |\Theta_\beta\rangle.
\end{eqnarray}

These expressions allow us to better understand the nature 
of Bell states: $U_+$ displaces both systems in position by 
the same amount while $V_-$ displaces them in momentum in 
opposite direction. As Bell states are common 
eigenstates of these operators, such states can be 
interpreted as corresponding to the eigenstates of the {\sl
total momentum} and {\sl relative position} (note that in the 
continuum limit  
$U_+=\exp(-i (P_1+P_2)\delta x/\hbar)$ and 
$V_-=\exp(i\delta p (Q_1-Q_2)/\hbar)$. In 
this sense, Bell states (\ref{bellstates}) are precisely the 
ones used by Einstein, Podolsky and Rosen \cite{EPR} 
to present their argument against completeness of quantum 
mechanics as a description of nature. 

Having this in mind, one expects the phase space
representation of Bell states to be rather simple. This is 
indeed the case:
\begin{eqnarray}
W_\beta(\alpha_1,\alpha_2)&=& {\mbox{Tr}}(|\Theta_\beta\rangle\langle
\Theta_\beta|\, \hat A(\alpha_1)\otimes \hat A(\alpha_2))\nonumber\\
&=& W_0(\alpha_1-2\beta, \alpha_2)\nonumber
\end{eqnarray}
where
\begin{eqnarray}
W_0(\alpha)&=&{\mbox{Tr}}(|\Theta_0\rangle\langle
\Theta_0|\, \hat A(\alpha_1)\otimes \hat A(\alpha_2))\nonumber\\
&=&{1\over (2N)^2} \delta_N(q_{\alpha_1}-q_{\alpha_2})
\delta_N(p_{\alpha_1}+p_{\alpha_2})\nonumber\\
&\,&\qquad \times 
(-1)^{(q_{\alpha_1}p_{\alpha_1}+q_{\alpha_2}p_{\alpha_2})/N}
\label{wbell0}
\end{eqnarray}
Thus, the Wigner function of $|\Theta_\beta\rangle$ 
is nonzero only when 
$q_{\alpha_1}-q_{\alpha_2}=2q_\beta$ and 
$p_{\alpha_1}+p_{\alpha_2}=2p_\beta$ (modulo $N$). The oscillations,
whose origin we described above for a simpler case (\ref{wigcompstate}), 
can also be interpreted 
as due to the interference with the mirror images created by the 
boundary conditions . Notice
that these are precisely the equations defining the manifold  
$L_{\beta}$. Thus, the projectors onto Bell state $|\Theta_\beta\rangle$
is simply the sum of phase space point operators over all 
points belonging to $L_\beta$. 

Before explicitly discussing teleportation it is useful 
to present some further results related to Bell states and 
their connection to phase space point operators. 
A complete basis of the space of operators on ${\cal H}^{(1,2)}$ 
can be constructed using Bell states: Thus, the operators 
$\hat B(\beta_1,\beta_2)=|\Theta_{\beta_1}\rangle\langle\Theta_{\beta_2}|$
form a complete orthogonal set (with $\beta_1,\beta_2\in G_N$). The 
change of basis between this set and phase space point operators
is 
\begin{equation}
\hat B(\beta_1,\beta_2)= 
(4N)^2\sum_{\alpha_1,\alpha_2\in G_N} K(\beta_1,\beta_2|
\alpha_1,\alpha_2) \, \hat A(\alpha_1,\alpha_2).\label{betatoalpha}
\end{equation}
where the coefficients $K(\beta_1,\beta_2|\alpha_1,\alpha_2)$ 
are, in general, complex numbers (when $\beta_1=\beta_2$ 
we have $K(\beta,\beta|\alpha_1,\alpha_2)=W_\beta(\alpha_1,\alpha_2)$).  
The precise form of this coefficients can be easily obtained but 
will not be needed here. The inverse of (\ref{betatoalpha}) 
is also useful and reads
\begin{equation}
\hat A(\alpha_1,\alpha_2)=\sum_{\beta_1,\beta_2} 
\tilde K(\alpha_1,\alpha_2|\beta_1,\beta_2) \hat 
B(\beta_1,\beta_2)
\label{alphatobeta}
\end{equation}
Simple relations between the coefficients of (\ref{betatoalpha}) and
(\ref{alphatobeta}) exist. In particular, one can show that
\begin{eqnarray}
\tilde K(\alpha_1,\alpha_2|\beta_1,\beta_2)&=& K(\beta_2,\beta_1|
\alpha_1,\alpha_2)\nonumber\\
&=& K^*(\beta_1,\beta_2 | \alpha_1,\alpha_2) \label{krelations}
\end{eqnarray}

Finally, we mention that Bell states satisfy the following
identity: 
\begin{equation}
{\mbox{Tr}_1}(\hat A(\alpha)\otimes {\mbox{I}}_2 
|\Theta_0\rangle\langle\Theta_0|)={1\over N} 
{\mbox{I}}_1\otimes \hat A^T(\alpha),\label{lastbell}
\end{equation}
where the transpose of the phase space point operator appears 
in the right hand side. 

\section{Teleportation in phase space}

We show here how the usual teleportation protocol can be 
described in phase space. We consider three identical subsystems
labeled by the integers $j=1,2,3$. The aim is to teleport the 
initial state of system $1$, which is characterized by an arbitrary
Wigner function $W(\alpha_1)$. For this we initially prepare
systems $2$ and $3$ in one of the Bell states (for simplicity 
we use $|\Theta_0\rangle_{2,3}$ as the initial state). Thus, the
initial density matrix of the combined three-partite system 
is 
\begin{eqnarray}
\rho_{1,2,3}&=& \rho_1\otimes |\Theta_0\rangle\langle\Theta_0|_{2,3}
\nonumber\\
&=&N^3\! \sum_{\alpha_1,\alpha_2,\alpha_3\in G_{2N}} 
W(\alpha_1)W_0(\alpha_2,\alpha_3)\nonumber\\
&\,&\qquad\qquad\times\, \hat A(\alpha_1)\otimes\hat A(\alpha_2)
\otimes \hat A(\alpha_3)\label{initialstate1}
\end{eqnarray}
where the Wigner function for the Bell state $|\Theta_\beta\rangle$ 
is given in (\ref{wbell0}). After preparing this initial state
the teleportation protocol proceeds as follows: First we perform a 
measurement of system $1$ and $2$ that projects them into the Bell
basis. Physically, as discussed above, this corresponds to a 
collective measurement that determines the total momentum 
$p_1+p_2$ and the relative coordinate $q_1-q_2$ for these two 
subsystems. After this measurement the state of the system
is projected into the resulting state $|\Theta_\beta\rangle_{1,2}$, 
where $\beta=(q_\beta,p_\beta)$ ($p_\beta$ and $q_\beta$ are 
the measured values of the total momentum and distance).  
To explicitly write down the resulting state, it is better to rewrite 
equation (\ref{initialstate1}) expressing the phase space point operator
$\hat A(\alpha_1)\otimes \hat A(\alpha_2)$ in terms of the Bell
operators $B(\beta_1,\beta_2)$ as in (\ref{betatoalpha}). Thus,
\begin{eqnarray}
\rho_{1,2,3}&=&(4N)^3\! \sum_{{\alpha_j\in G_{N}}\atop
{\beta_k\in G_N}}
W(\alpha_1)W_0(\alpha_2,\alpha_3)\tilde K(\alpha_1,\alpha_2|\beta_1,
\beta_2)\nonumber\\
&\,&\qquad\qquad\times\hat B(\beta_1,\beta_2)\otimes\hat A(\alpha_3)
\label{rho123}
\end{eqnarray}
From this equation it is obvious that, after the Bell measurement
of the first two subsystems, only the terms with $\beta_1=\beta_2=\beta$
in the above expression survive. Therefore, the state of the third 
subsystem is 
\begin{equation}
\rho'_3=\! 4N\sum_{\alpha_3\in G_{N}} W'(\alpha_3) 
\hat A(\alpha_3)\label{rho3}
\end{equation}
where the term multiplying the phase space point operator 
in (\ref{rho123}) can be identified as the new Wigner
function of the third system (up to a normalization constant). 
In turn, this Wigner function can be seen to be defined by an 
expression involving a sum that contains the initial Wigner 
function of the first system. Thus, this can be written as
\begin{equation}
W'(\alpha_3)= \sum_{\alpha_1\in G_{N}} Z(\alpha_3,\alpha_1) 
W(\alpha_1).
\end{equation}
The matrix $Z(\alpha_3,\alpha_1)$ simply tells us how to construct
the final Wigner function for the third system from the initial 
Wigner function of the first one. Again, the explicit form of
this matrix is read from (\ref{rho123}):
\begin{equation}
Z(\alpha_3,\alpha_1)=(2N)^4\sum_{\alpha_2\in G_{N}}
W_0(\alpha_3,\alpha_2) W_\beta(\alpha_1,\alpha_2). 
\label{zeta}
\end{equation}
The above expression is easy to evaluate since it 
contains a sum over half of the total phase space. Thus, we can 
use the relation (\ref{lastrelation}) to simplify it. 
Doing this (and using (\ref{lastbell})) one discovers 
that $Z(\alpha_3,\alpha_1)$ is just 
the trace of a product of two phase space point operators acting on 
the second subsystem which are evaluated at points $\alpha_1$ and 
$\alpha_3-2\beta$. Therefore, taking into account
the orthogonality conditions for phase space
point operators, one obtains
\begin{equation}
Z(\alpha_3,\alpha_1)= \delta_N(\alpha_3-\alpha_1-2\beta).\label{zfinal}
\end{equation}
This means that the state of the third subsystem has a Wigner function 
that is displaced in phase space by an amount $\beta$ with respect to the 
initial state of the first system. Therefore, to recover the initial 
state one must displace the third system in phase space by 
applying the evolution operator $\hat T(\beta)=
\hat T(q_\beta,p_\beta)$ (in fact, 
one can show that when the operator $\hat T(\beta)$ acts on the state
of the system, the Wigner function is simply shifted in phase 
space by the amounts $(2a,2b)$ the factors of $2$ being originated
in the fact that the grid has $2N\times 2N$ points \cite{MPS01}). 
Obviously, the recovery operation depends on $\beta$, which is 
revealed by the Bell measurement performed on the first two subsystems. 
In this way the final state of the third system is always 
identical to the unknown initial state of the first subsystem, 
which is the goal of the teleportation protocol.

\section{Measuring the Wigner Function}
 
The Wigner function, as mentioned above, is in one to one correspondence
with the quantum state of the system. Thus, experimentally determining
the value of $W(\alpha_1,\alpha_2)$ in every phase space point is
equivalent to completely determining the state of the system. 
Experimentally reconstructing the Wigner function has been the goal 
of a series of experiments, all dealing with continuous systems 
\cite{WignerMeas}. In general, these experiments are aimed at 
determining first marginal distribution of some observables and later
reconstructing the Wigner function by means of a Radon--like transform. 
However, it is interesting (and useful) to realize that there is 
no need to fully determine the quantum state to evaluate the Wigner
function at a given phase space point. Indeed, this was proposed originally 
by Davidovich and Lutterbach for the Wigner function of the state of 
the electromagnetic field stored in a superconducting cavity
\cite{DavidovichLut}. Using a minor variation of this method the 
direct measurement of the Wigner function at the origin of phase space
has been recently performed in a cavity QED experiment. 
More recently, this method was generalized and shown to be applicable
to the measurement of discrete (or continuous) 
Wigner functions of generic systems
in \cite{MPS01,NatureUS}. Here, we will describe how this tomographic
scheme can be generalized to directly measure the Wigner function 
we presented in this paper. The fact that 
this Wigner function can be measured in an efficient way (i.e., in 
a number of steps which scales polynomially with the dimensionality
of the system) may be important if, for example, one is able to 
relate interesting physical quantities (like some entanglement 
measures) with phse space observables (work is in progress 
in this direction).  

The efficient strategy to measure the Wigner
function of a composite system at (any) given phase space point is, 
as mentioned above, a direct  generalization of the idea 
originally proposed  in \cite{MPS01,NatureUS}
to measure the Wigner function of an $N$ dimensional system. 
The basic ingredient can be described
in terms of the following quantum algorithm. Consider a system 
initially prepared in a quantum state $\hat\rho$. We put 
this system in contact with an ancillary qbit prepared in the state 
$|0\rangle$. This ancillary qbit plays the role of a ``probe particle'' 
in a scattering--like experiment. The algorithm is: i) Apply 
an Hadamard transform to the ancillary qbit  
(where $H|0\rangle=(|0\rangle+|1\rangle)/\sqrt 2$, 
$H|1\rangle=(|0\rangle-|1\rangle)/\sqrt 2$), ii) Apply 
a ``controlled--$\hat M$'' operator (if the ancilla is in state
$|0\rangle$ this operator acts as the identity for the system but if 
the state of the ancilla is $|1\rangle$ it acts as the 
unitary operator $\hat M$ on the system), iii) Apply another 
Hadamard gate to the ancilla and finally perform a {\it weak} 
measurement on this qbit detecting its polarization (i.e., measuring 
the expectation values of Pauli operators 
$\sigma_z$ and $\sigma_y$). It is easy to show that the above
algorithm has the following remarkable property: 
\begin{equation}
\langle\sigma_z\rangle=Re(Tr(\hat M\hat\rho)), \quad
\langle\sigma_y\rangle=Im(Tr(\hat M\hat\rho)).
\end{equation}
Thus, the final polarization measurement of the ancillary 
qbit reveals a property determined both by the initial 
state $\hat\rho$ and the unitary operator $\hat M$.

In \cite{NatureUS} we discussed how to view this simple algorithm
as the basic tool to construct a rather general tomographer (and 
also a rather general spectrometer). In particular, 
we showed how to use it to 
measure the Wigner function of a simple system. Here, we 
show how to adapt it to measure the Wigner function
of the composite system we have been discussing so far. 
This can be done by applying the algorithm shown in Figure 1. 

\begin{figure}
\epsfxsize=7.6cm
\epsfbox{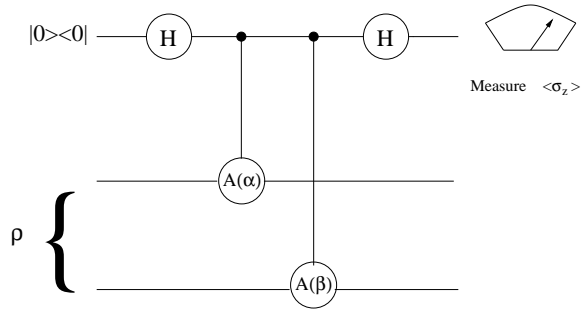}
\vspace {0.5cm}
\caption{Circuit for measuring $W(\alpha_1,\alpha_2)$, for 
a composite system. The controlled operations apply the 
operator $2N \hat A(\alpha_i)$ (which is unitary) 
conditioned on the state
of the ancilla qbit. The phase space operators parametrically depend
on the phase space point $\alpha_i$ and can be efficiently built 
as a simple sequence of displacements and reflections 
on each of the two subsystems. The measured polarization of 
the ancilla qbit is directly proportional to the Wigner function. }
\label{wcircuit}
\end{figure} 

From the previous discussion it is clear that the above algorithm
is such that by measuring the polarization of the ancillary 
qbit we determine the Wigner function. Indeed, this follows from
the identity
\begin{equation}
\langle \sigma_z\rangle=4N^2{\mbox{Tr}}(\rho \hat A(\alpha_1)\otimes
\hat A(\alpha_2))=4N^2W(\alpha_1,\alpha_2).
\end{equation}

As phase space point operators (\ref{disphase2})
are simply a product of displacement
operators (which implement addition of one, modulo $N$) and reflections
(which are the square of the Fourier transform) the network 
of Figure 1 can be implemented efficiently (i.e., it involves a 
number of elementary gates that grows polynomially with $\log(N)$). 

\section{Conclusion}

In this paper we used a hybrid approach to construct a Wigner function 
to represent quantum states of a composite system
in phase space. The function we defined 
has interesting features enabling us to study situations where 
entanglement between subsystems plays an important role. Thus, the
hybrid method captures some of the most useful properties of the 
Wigner functions defined by Wooters \cite{Wooters} and 
Leonhardt (and others)\cite{Leonhardt,MPS01}. For a bipartite
system this function depends upon two phase space coordinates 
$W(\alpha_1,\alpha_2)$. The phase space is a Cartesian product
of the phase spaces of the subsystems, as it is the case in Wotters 
proposal. However, each phase space grid has $2N\times 2N$ points, 
as suggested by Leonhardt and others. For separable states the 
Wigner function is, in general, a convex sum of products of 
independent functions for each 
subsystem. Thus, this Wigner function is a natural tool to study
entanglement between subsystems. In this paper we showed that 
basis of entangled states can be identified with non--separable
slices in the phase space (the basis formed by Bell states is 
one such example). 
We also showed that $W(\alpha,\alpha')$ is measurable by 
a simple scattering--like experiment where 
an ancillary particle successively interacts with the
two subsystems. 

\acknowledgments
JPP thanks L. Davidovich and Marcos Saraceno for useful discussions. 
He also thanks Cecilia Lopez for carefully reading the manuscript. 
This work was partially supported with grants from 
Ubacyt, Anpcyt and Fundaci\'on Antorchas. JPP is a fellow 
of CONICET.


\begin{references}

\bibitem{Teleport1} C. H. Bennett, G. Brassard, C. Cr\'epeau, 
R. Jozsa, A. Peres and W. K. Wooters, {\it Phys. Rev. Lett.} 
{\bf 70}, 1895 (1993).

\bibitem{TeleExperiment} D. Bouwmeester, J-W. Pan, K. Mattle, M. Eibl, 
H. Weinfurter and A. Zeilinger, {\it Nature} {\bf 390} 575 (1997).

\bibitem{Chuang-Nielsen} {\it "Quantum Information and Computation"}, 
I. Chuang and M. Nielsen (2000), Cambridge University Press.

\bibitem{GottChuang} D. Gottesman and I. Chuang, {\it Nature} {\bf 402} 
390 (1999). 

\bibitem{Vaidman} L. Vaidman, {\it Phys. Rev.} {\bf A49}, 1473 (1993).

\bibitem{Braunstein} S.L. Braunstein and H. J. Kimble, 
{\it Phys. Rev. Lett.} {\bf 80} 869 (1999).

\bibitem{Kimbleexp} A. Furusawa {\it et al}, 
Science {\bf 282} (1998) 706. 

\bibitem{Wigner} M. Hillery, R.F. O'Connell, M.O. Scully, E. P. 
Wigner, {\it Phys. Rep.}{\bf 106} 121 (1984).

\bibitem{Buzek} M. Koniorczyk, V. Buzek and J. Jansky, 
{\it Phys. Rev.} {\bf A64} (2001) 034301. 

\bibitem{Wooters} W. K. Wooters, Ann. Phys. NY {\bf 176} (1987), 1

\bibitem{Leonhardt} U.Leonhardt, Phys. Rev. Lett. {\bf 74} (1995) 4101; 
U.Leonhardt, Pys. Rev. A {\bf 53} (1996) 2998.

\bibitem{Hannay} J. H. Hannay, M. V. Berry, 
{\it Physica} {\bf 1D} (1980) 267.

\bibitem{Rivas} A. Rivas, A. M. Ozorio de Almeida, 
Ann.Phys. {\bf 276} (1999), 123

\bibitem{Bouzouina} A. Bouzouina, S. De Bievre, 
Comm. Math. Phys. {\bf 178} (1996)83

\bibitem{WignerUS} P. Bianucci, C. Miquel, J. P. Paz and M. Saraceno,
{\it Discrete Wigner functions and the phase space representation of
a quantum computer} quant-ph/0105091.

\bibitem{MPS01} C. Miquel, J. P. Paz and M. Saraceno, 
{\it Quantum computers in phase space}, (2001), submitted to PRA. 

\bibitem{Schwinger} J. Schwinger, {\it Proc. Nat. Acad, Sci.}{\bf 46} 
(1960), 570, 893. 

\bibitem{EPR} A. Einstein, B. Podolsky and N. Rosen, {\it Phys. Rev} 
{\bf 47} (1935) 777.

\bibitem{WignerMeas} T. J. Dunn {\it et al.},
Phys. Rev. Lett. {\bf 74} (1994) 884;
D. Leibfried {\it et al.}, 
Phys. Rev. Lett. {\bf 77} (1996) 4281;
see also Physics Today {\bf 51} no. 4 (1998) 22;
L. Lvovsky {\it et al.}, 
Phys. Rev. Lett. {\bf 8705} (2001) 050402.

\bibitem{DavidovichLut} L. G. Lutterbach and L.
Davidovich, 
Phys. Rev. Lett. {\bf 78} (1997) 2547; Optics Express {\bf 3} (1998) 147.

\bibitem{Harocheetal} G. Nogues et al,
Phys. Rev. {\bf A62} (2000) 054101

\bibitem{NatureUS} C. Miquel, J.P. Paz, M. Saraceno, E. Knill, 
R. Laflamme, C. Negrevergne (2001), quant-ph/0109072. 

\end{references}
\end{document}